\documentclass[12pt,preprint]{aastex}

\usepackage{graphicx}
\usepackage{comment}
\usepackage{amsmath}
\usepackage{epstopdf}

\def\beq{\begin{equation}} 
\def\eeq{\end{equation}} 
\def\beqar{\begin{eqnarray}} 
\def\eeqar{\end{eqnarray}}

\def\pfrac#1#2{\left( \frac{#1}{#2} \right)}

\def\msol{M_\odot}

\def\fradio{f_{\rm radio}}
\def\fradval{10\%}

\def\la{\mathrel{\mathpalette\fun <}}
\def\ga{\mathrel{\mathpalette\fun >}}
\def\fun#1#2{\lower3.6pt\vbox{\baselineskip0pt\lineskip.9pt
  \ialign{$\mathsurround=0pt#1\hfil##\hfil$\crcr#2\crcr\sim\crcr}}}

\def\pfrac#1#2{\left( \frac{#1}{#2} \right)}

\begin{document} 

\title{Radio Supernovae in the Great Survey Era}

\author{Amy Lien, Nachiketa Chakraborty, Brian D. Fields\altaffilmark{1}, 
Athol Kemball}

\affil{Department of Astronomy, University of Illinois, Urbana, IL}

\altaffiltext{1}{Department of Physics, University of Illinois, Urbana, IL}

\begin{abstract}

Radio properties of supernova outbursts 
remain poorly understood despite 
longstanding campaigns following events
discovered at other wavelengths.
After $\sim 30$ years of observations,
only $\sim 50$ supernovae have been detected at radio wavelengths, 
none of which are Type Ia.
Even the most radio-loud events 
are $\sim 10^4$ fainter in the radio than in the optical;
to date, such intrinsically dim objects have only been
visible in the very local universe.
The detection and study of radio supernovae (RSNe)
will be fundamentally altered and
dramatically improved
as the next generation of 
radio telescopes comes online,
including EVLA, ASKAP, and MeerKAT, and
culminating in the Square Kilometer Array (SKA);
the latter should be 
$\ga 50$ times more sensitive than present facilities.
SKA can
repeatedly scan large ($\ga 1 \, \rm deg^2$) areas of the sky,
and thus will discover RSNe and other transient sources
in a new, automatic, untargeted, and unbiased way.
We estimate
SKA will be able to 
detect core-collapse RSNe out to redshift $z \sim 5$,
with an all-redshift rate
$\sim 620 \ {\rm events \ yr^{-1} \ deg^{-2}}$,
assuming a survey sensitivity of 50 nJy
and radio lightcurves like those of SN 1993J.
Hence SKA should provide a complete
core-collapse RSN sample 
that is sufficient for
statistical studies of radio properties of core-collapse supernovae.
EVLA should find $\sim 160 \ {\rm events \ yr^{-1} \ deg^{-2}}$
out to redshift $z \sim 3$,
and other SKA precursors should have similar detection rates.
We also provided recommendations of the survey strategy
to maximize the RSN detections of SKA.  
This new radio core-collapse supernovae sample will complement 
the detections from the optical searches, such as the LSST,
and together provide crucial information on massive star evolution,
supernova physics, and 
the circumstellar medium, out to high redshift.
Additionally, SKA may yield the first radio Type Ia detection via
follow-up of nearby events discovered  at other wavelengths.

\end{abstract}

\noindent

\section{Introduction}

\label{sect:intro}

Supernovae are among the most energetic phenomena
in the universe, and are central to
cosmology and astrophysics.
For example, 
core-collapse supernovae are 
explosions arise from the death of massive stars
and hence are closely
related to the cosmic star-formation rate
and to massive-star evolution, 
and are responsible for the energy and baryonic feedback of the environment
\citep{madau}.
Type Ia supernovae show a uniform properties in their lightcurves 
and play a crucial role as
cosmic ``standardizable candles'' \citep{Phillips93,Riess96}.

Our knowledge of the {\em optical} properties of supernovae,
is increasing rapidly with the advent of prototype ``synoptic''--i.e.,
repeated scan--sky surveys,
such as SDSS-II \citep{Frieman08,Sako08} and SNLS \citep{Bazin09,PD10}.
These campaigns
are precursors to the coming ``Great Survey'' 
era in which synoptic surveys
will be conducted routinely over very large regions of sky,
e.g.,~LSST \citep{lsst,lsstsb} and Pan-STARRS \citep{pan-starrs-sne}.  
The number of detected supernovae
will increase by several orders of magnitude in 
this era
\citep{lsstsb,lf}.

In contrast to this wealth of optical information,
properties of supernovae in the radio
remain poorly understood, fundamentally due to observational limitations. 
Radio supernovae (RSNe) have primarily 
been discovered by follow-up observations 
of optical outbursts,
and only very rarely by accident.
To date, only $\sim 50$ core-collapse outbursts
have radio detections, and no Type Ia
explosion has ever been detected in the radio
\citep{Weiler04, Panagia06}.
The core-collapse subtype Ibc has been a focus of recent
study in the radio, because some Type Ibc events 
are associated with
long Gamma-Ray Bursts (GRBs) 
\citep{Galama98, Kulkarni98, Soderberg07, Berger03}.
 
Current radio interferometers are 
scheduled primarily around targeted observations proposed by 
individual principal investigators. 
This stands in contrast to future radio interferometers 
planned for the coming ``Great Survey'' era.
These include the Square Kilometer Array 
(SKA\footnote{http://www.ska-telescope.org}) 
and its precursor prototype arrays 
(for example, ASKAP\footnote{http://www.atnf.csiro.au/projects/askap} 
and MeerKAT\footnote{http://www.ska.ac.za/meerkat}). 
These telescopes will operate primarily 
as wide-field survey instruments focusing on 
several key science projects \citep{Carilli04}. 
As synoptic telescopes, 
they will be far better suited to study 
all classes of transient and time-variable radio sources, including RSNe.
\citet{Gal-Yam06} already pointed out the power of synoptic radio surveys 
for detecting radio transients of various types, 
including supernovae and GRBs, in an unbiased way. 
Here we quantify the prospects for RSNe.

In this paper we explore this fundamentally new mode of
{\em untargeted} RSN discovery and study.
We adopt a forward-looking perspective,
and consider the new science enabled by
RSNe observations in an era in which the full SKA is operational.
Our focus is
mainly on core-collapse supernovae, the type for which some radio
detections
already exist.
However, we will also discuss the possibility of 
Type Ia radio discovery based on
current detection limits. 
We will first summarize 
current knowledge of radio core-collapse supernovae 
(\S \ref{sect:properties}), 
and the expected sensitivity of SKA 
(\S \ref{sect:sensitivity}). 
Using this information, we forecast
the radio core-collapse supernovae harvest of SKA (\S \ref{sect:formalism}),
and consider optimal survey 
strategies (\S \ref{sect:recommend}).
We conclude by anticipating the RSN science payoff
in this new era (\S \ref{sect:science}). 
We adopt a standard flat $\Lambda$CDM 
model with $\Omega_{\rm m} = 0.274$, $\Omega_{\Lambda}=0.726$,
and $H_0 = 70.5 \, \rm km \, s^{-1} \, Mpc^{-1}$
\citep{wmap5} throughout.

\section{Radio Properties of Supernovae}
\label{sect:properties}

Several key properties of RSNe have been established,
as a result of the longstanding leadership of the
NRL-STScI group \citep[recently reviewed in][]{Weiler09,Stockdale07,Panagia06}
and of the CfA group and others \citep[summarized in][]{Soderberg07,Berger03}.
We summarize these general RSN characteristics, which we will
use to forecast the RSN discovery
potential of synoptic radio surveys.  

\subsection{Radio Core-Collapse Supernovae}
\label{sect:radioCC}

Observed core-collapse RSNe
have luminosities spanning 
$\nu L_\nu \sim 10^{33} - 10^{38} \ {\rm erg \ s^{-1}}$
at 5 GHz,
and thus are $\ga 10^4$ times
less luminous in the radio than in the optical.
Their intrinsic faintness has prevented
RSN detection in all but the most local universe.
Even within a particular core-collapse subtype,
radio luminosities 
and lightcurves are highly
diverse, e.g., two optically similar Type Ic events
might be radio bright in one case and 
undetectable in the other \citep{1997X, 1997ei, rsnweb}
\footnote{New Radio Supernova Results \citep{rsnweb} 
are available online at: http://rsdwww.nrl.navy.mil/7213/weiler/sne-home.html}.
Additionally, core-collapse RSNe spectral shapes strongly evolve with time;
lightcurves peak over days to months depending
on the frequency.
RSN emission can be understood
in terms of interactions between the blast, 
ambient relativistic electrons, and the circumstellar medium
\citep{Chevalier82a, Chevalier82b, Chevalier98}. 

To model
RSN emission as a function of
frequency and time, we adopt the semi-empirical form
derived by \citet{Chevalier82a} and extended 
in \citet{Weiler02},
\beq
L(t, \nu) = L_1 \pfrac{\nu}{5 \, \rm GHz}^{\alpha} \, 
         \pfrac{t}{1 \, \rm day}^{\beta} \,
         e^{-\tau_{\rm external}} \,
         \pfrac{1-e^{-\tau_{\rm CSM_{\rm clumps}}}}{\tau_{\rm CSM_{\rm clumps}}} \,
         \pfrac{1-e^{-\tau_{\rm internal}}}{\tau_{\rm internal}} \ \ . 
\eeq
We follow the notation of \citet{Weiler02}.
$L(t, \nu)$ is the supernova luminosity at 
frequency $\nu$ and time $t$ after the explosion.
Optical depths from material both outside 
($\tau_{\rm external}$, $\tau_{\rm CSM_{\rm clumps}}$) 
and inside ($\tau_{\rm internal}$) the blast-wave front 
are taken into account \citep[see][]{Weiler02}.

Parameters embedded in
each optical depth term are those for SN 1993J,
one of the best studied RSNe 
\citep{Weiler07}.
Radio emission from SN 1993J is dominated by 
the clumped-circumstellar-medium (clump-CSM) term, and hence
\beq
\label{eq:lum}
L(t, \nu) \sim \frac{1-e^{-\tau_{\rm CSM_{\rm clumps}}}}{\tau_{\rm CSM_{\rm clumps}}},
\eeq
where $\tau_{\rm CSM_{\rm clumps}} = 4.6 \times 10^5 
\ (\frac{\nu}{\rm 5 \ GHz})^{-2.1} \ 
(\frac{t}{\rm 1 \ day})^{-2.83}$,
for SN 1993J.
At small $t$, $\tau_{\rm CSM_{\rm clumps}}$ is large and 
$L(t, \nu) \sim 1/\tau_{\rm CSM_{\rm clumps}} \propto \nu^{2.1} \ t^{2.83}$,
so luminosity grows as a power law at 
early times. With all optical depth parameters fit to
SN 1993J, the peak luminosity is controlled by the prefactor $L_1$.

\begin{figure}[!h]
\begin{center}
\includegraphics[width=0.6\textwidth]{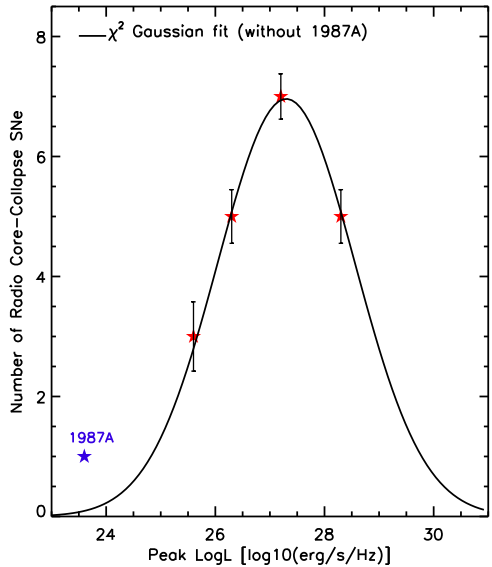}
\end{center}
\caption{
Radio luminosity function (not-normalized) at 5 GHz 
of core-collapse supernovae showing
core-collapse supernovae count as 
a function of $\log_{10}(L)$, where L is the peak radio luminosity.
Data are binned to 
$\Delta \log_{10}(L) = 1$. 
The black curve shows the $\chi^2$-fitted Gaussian 
to the underlying data (red stars).
}
\label{fig:radio_lum}
\end{figure}

Our main focus will be on RSN discovery, 
and thus it is most important to capture the wide
variety of peak radio luminosities, which correspond in our model
to a broad distribution for $L_1$.
Figure~\ref{fig:radio_lum} shows a crude 
luminosity function (not-normalized) based on
the sample of 20 core-collapse supernovae 
(15 Type II and 5 Type Ibc)
that have a published peak luminosity 
at 5 GHz \citep{Weiler04, Stockdale03, Stockdale07, Papenkova01, rsnweb, 
Baklanov05, Pooley02}.
We use 5 GHz data to construct our luminosity function
because the most RSNe have been observed at this frequency.
However, our predictions will span a range of frequencies,
based on this luminosity function and eq.~(\ref{eq:lum}).
The data are divided into four luminosity bins of size 
$\Delta \log_{10}(L) = 1$.
The black curve in Fig.~\ref{fig:radio_lum} is the best-fit
Gaussian, with average luminosity
$\log_{10}(L_{\rm avg}/\rm erg \, s^{-1} \, Hz^{-1}) = 27.3$, a 
standard deviation $\sigma = 1.25$, and $\chi^2 = 0.18$.  
SN 1987A is marked in Fig.~\ref{fig:radio_lum},
but was not used in the fit 
to avoid possible bias due to its
uncommonly low luminosity.
The fitted luminosity function might be biased towards
the brighter end,
because of the current survey sensitivity 
and the small and incomplete nature of the sample.

\subsection{Radio Type Ia Supernovae}
\label{sect:radioIa}

All searches to date have failed to
detect radio emission from Type Ia supernovae.
\citet{Panagia06} reported the radio upper limits of 27
Type Ia supernovae from more than two decades of observations
by the Very Large Array (VLA). The weakest
limit on a Type Ia event is 
$4.2 \times 10^{26} \ \rm erg \ s^{-1} \ Hz^{-1}$
at 1.5 GHz for SN 1987N,
which is around one order of magnitude lower than
the average luminosity of radio-detected core-collapse
supernovae (see \S \ref{sect:radioCC}).
The strongest limit on Type Ia radio emission
is even tighter, 
$8.1 \times 10^{24} \ \rm erg \ s^{-1} \ Hz^{-1}$ at 8.3 GHz
for SN 1989B.
Additionally, the $z \sim 0$ cosmic Type Ia supernova rate is
around 1/4.5 of the core-collapse supernova rate \citep{Bazin09}.
The intrinsic faintness in radio
and their smaller rate make detecting Type Ia in radio observations
especially hard.

\section{Next-Generation Radio Telescopes:  Expected Sensitivity}
\label{sect:sensitivity}

Radio detections of supernovae to date have been restricted both by the limiting sensitivity of contemporary radio interferometers and the need for
dedicated telescope time for transient followup.  This situation will
change drastically with SKA's unprecedented sensitivity and particularly
by its ability to repeatedly scan large regions of the sky at this great depth.

Current SKA specifications adopt 
a target sensitivity parameter 
$A_{\rm eff}/T_{\rm sys} = 10^4 \ \rm m^2 \ K^{-1}$ 
at observing frequencies in the low several GHz,
including $z = 0$ HI at 1.4 GHz.
$A_{\rm eff}$ is the effective aperture, and $T_{\rm sys}$ is
the system temperature.  We will adopt this value of 
$A_{\rm eff}/T_{\rm sys}$, which yields a 1-$\sigma$ rms thermal noise limit 
in total intensity of
\beq
\sigma_I=0.15 \ \rm \mu Jy \ (\Delta \nu/{\rm GHz})^{-1/2} \  
   ({\delta t}/{\rm hr})^{-1/2}, 
\label{eq:rms}
\eeq
for a bandwidth $\Delta \nu$ 
and observation duration $\delta t$. 
The SKA will therefore reach a thermal noise limit of 
several nJy in deep continuum integrations ($\delta t \sim 1000$ hr) 
\citep{DRM09}
\footnote{http://www.skatelescope.org/PDF/DRM\_v1.0.pdf}. 
We define the associated survey sensitivity $S_{\rm min}$ 
(the minimum flux density threshold) as $S_{\rm min}=3\sigma_I$. 
In common with other radio
interferometers, 
SKA will accumulate sensitivity 
in targeted deep fields, 
including transient-monitoring fields, 
by accumulating integration time
over multiple individual observing tracks. 
We therefore will adopt a fiducial SKA supernova sensitivity of
$S_{\rm min} = 50 \ \rm nJy$ in 100 hours of observation,
but we will show how our results are sensitive to other
choices of $S_{\rm min}$.

It is anticipated that transient fields will be 
revisited with a cadence appropriate to the variability timescales
under study and that interferometric inverse imaging methods 
will include source models with time variability. 
Survey optimization for
interferometric transient detection is an active area of 
current SKA research. The technical details are beyond 
the intent and scope of this paper,
but will be influenced by science goals for 
transient source study in general, including pulsars, 
GRBs, and supernovae (as considered in this
paper), as well as the as-yet undiscovered transient population.

\section{Radio Supernovae for SKA}
\label{sect:formalism}

With its unprecedented sensitivity, 
SKA will be capable of synoptic search for 
core-collapse RSNe and open new possibilities
in radio astronomy. 
In this section, we predict the RSN detections of SKA
based on current knowledge to demonstrate how the RSN survey
can be done. 

\subsection{Core-Collapse Supernovae}
\label{sect:cc}

The detection rate 
$\Gamma_{\rm detect} = dN_{\rm SN}/(dt_{\rm obs} \, d\Omega \, dz)$ 
for a given RSN survey is
\beq
\label{eq:r_detect}
\Gamma_{\rm detect} = f_{\rm survey}\ f_{\rm radio}
                 \ f_{\rm ISM} \  \Gamma_{\rm SN} \ \ ,
\eeq
and is set by several observability factors $f$ 
that modulate the total rate of all supernovae 
\beq
\Gamma_{\rm SN}(z) = 
\frac{dN_{\rm SN}}{dV_{\rm comov} \, dt_{\rm em}} \ 
\frac{dt_{\rm em}}{dt_{\rm obs}} 
\ \frac{dV_{\rm comov}}{d\Omega \, dz}
 = {\cal R}_{\rm SN}(z) \,  r^2_{\rm comov}(z) \, c \left|\frac{dt}{dz}\right|
\eeq 
within the cosmic volume out to
redshift $z$ \citep{madau,lf}.

We see that the total cosmic supernova rate $\Gamma_{\rm SN}$
depends on cosmology via the volume element and the time dilation
terms. Because $\Lambda$CDM cosmological parameters are now known
to high precision,
these factors have a negligible error compared to the
other ingredients in the calculation.
The other factor in $\Gamma_{\rm SN}$
is the cosmic core-collapse supernova rate
density 
${\cal R}_{\rm SN}(z) = dN_{\rm SN}/(dV_{\rm comov} \, dt_{\rm emit})$.
Some direct measurements of 
this rate now exist out to $z \sim 1$, 
but the uncertainties remain large
\citep{Cappellaro99,Dahlen04,Cappellaro05,
Hopkins06,Botticella08,Dahlen08,Kistler08,Bazin09,Smartt09,
Dahlen10,Li11c,Horiuchi11}.
However, core-collapse events are short-lived,
and so the cosmic core-collapse rate is proportional to the
cosmic star-formation rate $\dot{\rho}_\star$,
which is much better-determined and extends to much higher redshifts.
We thus derive ${\cal R}_{\rm SN}$ from
the recent \citet{Horiuchi09}
fit to the cosmic star-formation rate.
The proportionality follows from the choice of 
initial mass function;
we apply 
the Salpeter initial mass function \citep{Salpeter55}
and assume the mass range of core-collapse SNe progenitors to be
$8\msol - 50\msol$; this gives 
${\cal R}_{\rm SN} = (0.007 \, \msol^{-1}) \, \dot{\rho}_\star$

Several effects reduce the total rate $\Gamma_{\rm SN}$ to the observed rate
$\Gamma_{\rm detect}$ in eq.~(\ref{eq:r_detect}).
Due to finite survey sensitivity, only 
a fraction $f_{\rm survey}$ of events are bright enough to detect, and only some fraction $f_{\rm radio}$ of supernovae 
will emit in the radio.
We neglect interstellar extinction and assume $f_{\rm ISM} \sim 1$ at the radio wavelengths considered. 

The term $\fradio$ in eq.~(\ref{eq:r_detect}) contains the greatest
uncertainty due to the relatively small sample of RSNe observed
to date, and the unavoidable incompleteness of the sample
(K. Weiler, private communication 2010). The
only published fraction available is for Type Ibc supernovae.
Using VLA for radio follow-up, \citet{Berger03}
suggests that $f_{\rm radio,Ibc} \sim 12\%$
after surveying 33 optically-detected Type Ibc supernovae.
For the purpose of demonstration,
we will adopt $\fradio = 10\%$ for the calculations presented
in this paper, which we believe is rather conservative.

An order-of-magnitude calculation
provides a useful estimate of the expected core-collapse
RSN rate.
As discussed in \S \ref{sect:sensitivity},
we adopt a fiducial SKA sensitivity of
$S_{\rm min} = 50 \ \rm nJy$.
Hence SKA will be able to detect supernovae with average
radio luminosity ($L \sim 10^{27} \ \rm erg \ s^{-1} \ Hz^{-1}$)
to a distance 
$D_L = \sqrt{L/4 \pi S_{\rm min}} \sim 4 \ \rm Gpc$,
which for a $\Lambda$CDM cosmology corresponds to 
$z \sim 1$.
This will give a detectable volume of
$V_{\rm detect} \sim (4/3) \pi D^3_L \sim 2.85 \times 10^{11} \ \rm Mpc^3$.
Observations show that the core-collapse supernova rate
$R_{\rm SN} \sim 10^{-3} \ \rm yr^{-1} \ Mpc^{-3}$ at $z \sim 1$
\citep{Dahlen08,Dahlen10}.
Assuming the fraction of the total core-collapse supernovae that display
the adopted average radio luminosity to be $f_{\rm radio} \sim 10\%$
\citep{Berger03},
the all-sky detection rate $ dN_{\rm SN}/dt \sim R_{\rm SN} \times f_{\rm radio} \times V_{\rm detect} \sim 2.85 \times 10^7 \ \rm yr^{-1}$.
This corresponds to a areal detection rate 
$dN_{\rm SN}/(dt \, d\Omega) \sim 700 \ \rm yr^{-1} \ deg^{-2}$.
As we now see, a more careful calculation confirms this estimate.

\begin{figure}[!h]
\begin{center}
\includegraphics[width=0.85\textwidth]{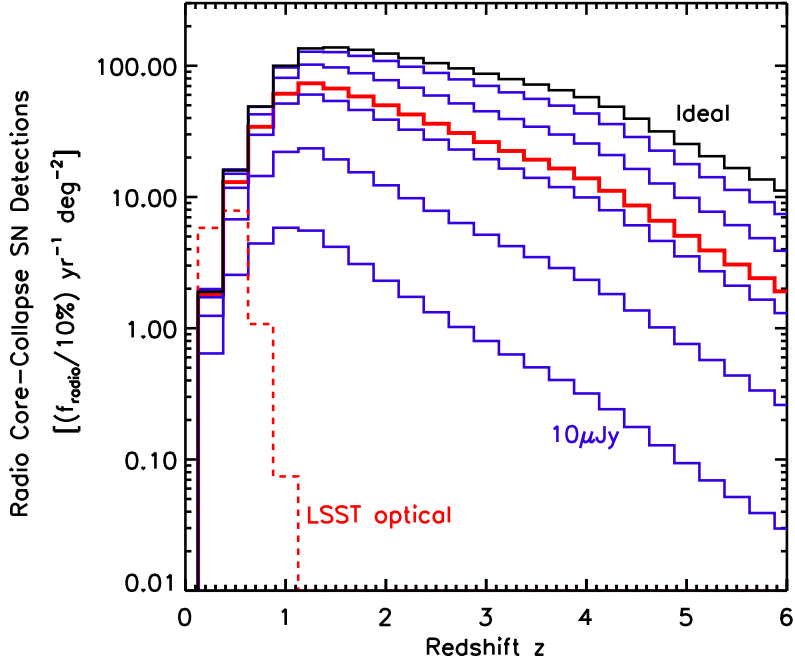}
\end{center}
\caption{
Estimated radio core-collapse supernova detection rate
as a function of redshift 
at 1.4 GHz, assuming $\fradio = \fradval$. Predictions 
are shown for different
survey sensitivities: 
$S_{\rm min}$ = $\{$10 $\mu$Jy (blue), 1 $\mu$Jy (blue),  100 nJy (blue), 50 nJy (thick red), 10 nJy (blue), 1 nJy (blue)$\}$
from bottom to top solid curves, respectively.
We adopt 50 nJy as our benchmark sensitivity hereafter.
For comparison, the red-dashed curve 
shows the LSST optical supernova detection rate
per year per $\rm deg^2$ \citep{lf}.
Also, the top solid curve (black) plots the 
ideal core-collapse RSN rate for comparison.
}
\label{fig:radio_obs}
\end{figure}

\begin{figure}[!h]
\begin{center}
\includegraphics[width=0.6\textwidth]{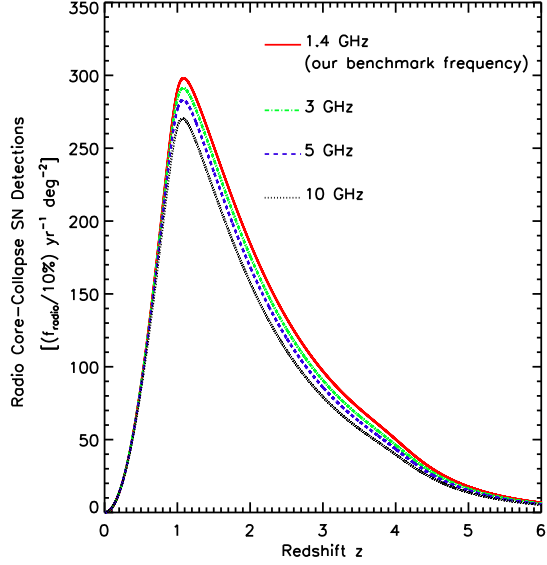}
\end{center}
\caption{
Estimated radio core-collapse supernova detection rate
as a function of redshift for different frequency bands,
for $\fradio = \fradval$, 
and an 
adopted survey sensitivity $S_{\rm min} = 50$ nJy.
}
\label{fig:radio_obs_band}
\end{figure}

\begin{figure}[!h]
\includegraphics[width=0.56\textwidth]{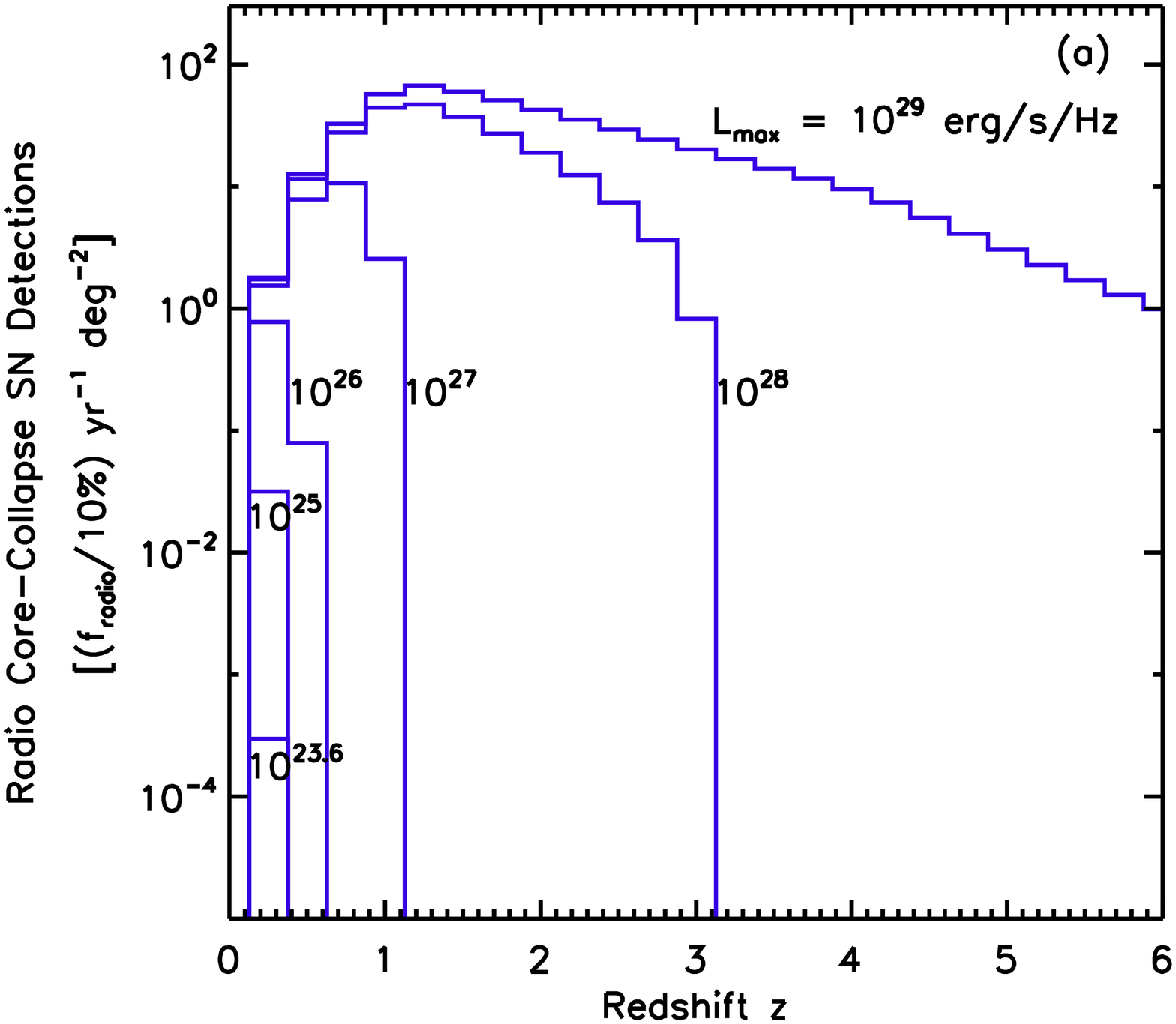}
\includegraphics[width=0.56\textwidth]{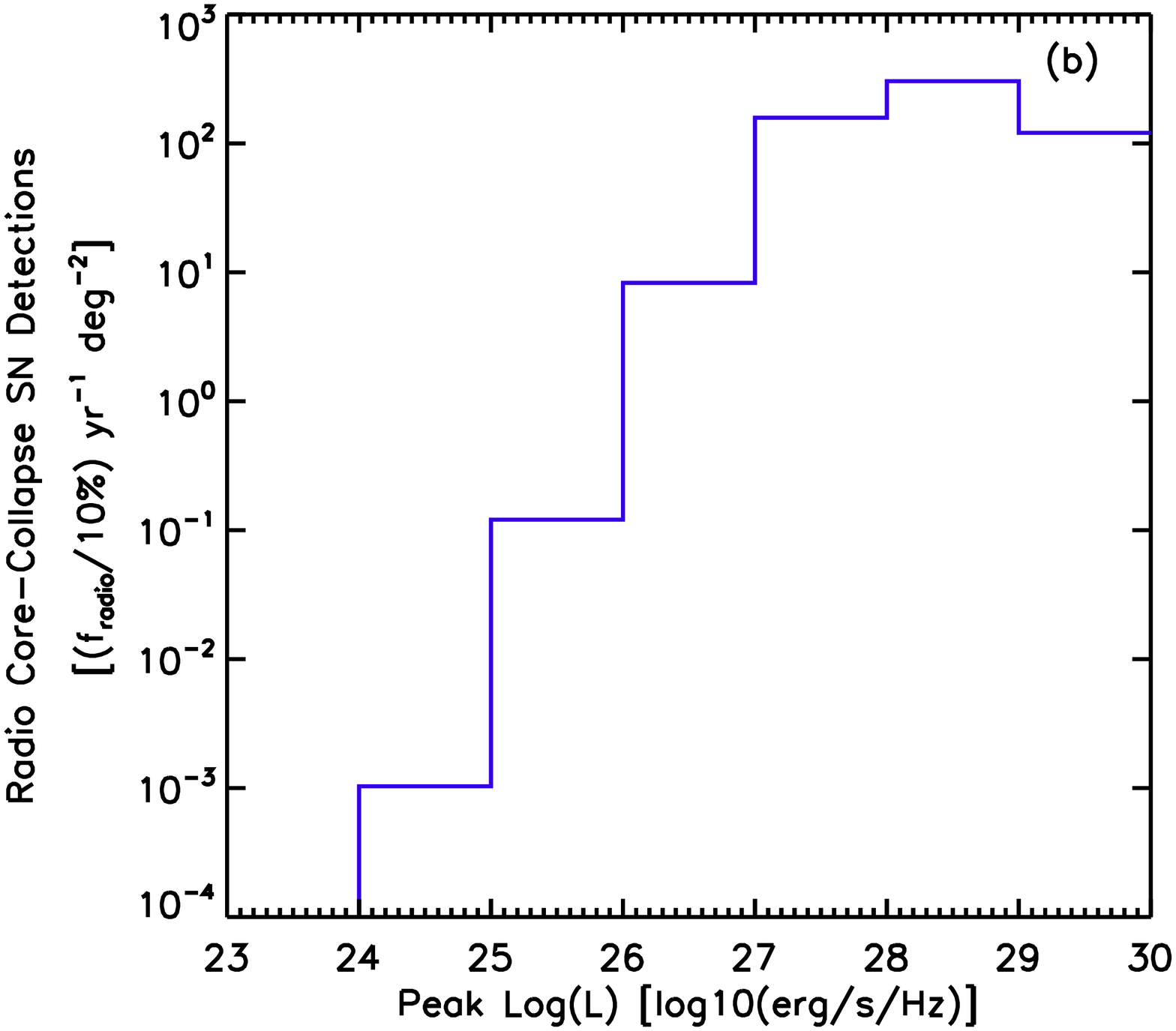}
\caption{
Core-collapse detection sensitivity to supernova radio luminosity,
at 1.4 GHz, and for survey sensitivity $S_{\rm min} = 50$ nJy.
(a){\em Left Panel}:
Supernova distribution over redshift,
for different cutoffs for the luminosity function.
(b){\em Right Panel}:
Supernova distribution in luminosity bins,
integrated over all redshifts.
}
\label{fig:radio_obs_lum}
\end{figure}

A careful prediction involves detailed calculation of 
$f_{\rm survey}(z)$.
The fraction $f_{\rm survey}(z)$ of observable radio-emitting events
depends on adopted survey sensitivity, and 
on the normalized supernova luminosity function $\Phi_{\rm 5GHz}(\log L)$,
which is measured at a peak luminosity at 5 GHz 
(derived in \S~\ref{sect:properties}).
In this paper we will only consider whether 
a supernova is detectable at its peak luminosity at 
each corresponding frequency.
The peak radio luminosity should be reached earlier at 
higher frequencies 
because of preferential absorption at lower frequencies
\citep{Weiler02}.
At different redshift, the peak flux density $S^{\rm peak}_{\rm min}$ 
in the observed frequency $\nu$ can be tied
to the corresponding luminosity threshold $L^{\rm peak}_{\rm min}$ by
\beq
\label{eq:Lmin}
L^{\rm peak}_{\rm min}(z; \nu_{\rm emit}) = \frac{4 \pi D^2_L(z)}{(1+z)} S^{\rm peak}_{\rm min}(\nu_{\rm obs}),
\eeq
where the luminosity distance is
$D_L(z) 
    = (1+z) \ c/H_0 \ \int^z_0
        dz^\prime \ [\Omega_{\rm m} (1+z^\prime)^3 + \Omega_{\Lambda}]^{-1/2}$ .
However, because the luminosity function we used is based on the peak luminosity
at 5 GHz, we must find the corresponding luminosity threshold
at this frequency by applying corrections based on the radio spectrum, 
\beq
L^{\rm peak}_{\rm min, 5GHz} = L^{\rm peak}_{\rm min} \ 
               \frac{\int_{\rm 5 GHz \ band} S^{\rm peak}(\nu_{\rm em}) \ d\nu_{\rm em}}
            {\int_{\rm obs \ band} S^{\rm peak}[ (1+z)\nu_{\rm obs} ]  \ d\nu_{\rm obs}}.
\eeq
The detectable fraction resulting from  survey sensitivity can therefore 
be estimated as 
\beq
f_{\rm survey}  = \int_{\log L^{\rm peak}_{\rm min, 5GHz}} \Phi_{\rm 5GHz}(\log L) \, d \log L.
\eeq 

Figure.~\ref{fig:radio_obs} plots the results 
of our predicted core-collapse RSN detection rate 
for different target survey sensitivities, $S_{\rm min}$.
We adopt a benchmark frequency of 1.4 GHz
because this will be one of 
the first major bands SKA deploys to observe
neutral hydrogen.
The related instantaneous field-of-view 
at 1.4 GHz
of current SKA designs based on dish reflectors is 
approximately 1 $\rm deg^2$, 
which we adopt.
Fig.~\ref{fig:radio_obs}
plots the ideal core-collapse supernova rate 
for comparison
(assuming infinite sensitivity but $\fradio = \fradval$).
One can see that the detection rate at 1 nJy
is very close to 
the ideal rate in the universe.

Results for our
fiducial SKA flux limit $S_{\rm min}$ = 50 nJy
are highlighted in
Fig.~\ref{fig:radio_obs}.
At this sensitivity, we see that we can expect that
radio supernovae will be discovered 
(event rates $> 5 \ \rm RSN \, yr^{-1} \, deg^{-2}$)
over the enormous redshift range $z \simeq 0.5$ to 5.
The total rate of RSNe expected in this entire redshift range
is 
\beq
\frac{dN_{\rm SN}}{dt \, d\Omega}(>50 \ {\rm nJy})
 \approx 620 \ \rm RSNe \, yr^{-1} \, deg^{-2} \ \ ,
\eeq
in good agreement with our above order-of-magnitude estimate.
This sample size is large enough to be statistically useful
and to allow for examination of the redshift history of RSNe.
Moreover, out to $z \sim 1$, SKA will detect almost all cosmic RSNe 
in the field of view, while at higher redshift the detections 
still comprise $> 10\%$ of the underlying ideal cosmic rate.
For comparison, we also see that LSST will detect optical supernovae
out to $z \la 1$.  Thus SKA will be complementary to
LSST as a unique tool for cosmic supernova discovery.

Figure.~\ref{fig:radio_obs_band} shows 
how core-collapse RSN detections 
vary for different observing frequencies,
fixing a common 
survey sensitivity $S_{\rm min}$ = 50 nJy and 
bandwidth $\Delta \nu = 1$ GHz.
Results show similar numbers of detections 
at different bands, which is 
caused by a relatively flat spectrum shape 
at peak luminosities.
Because SKA will be able to detect core-collapse RSNe 
out to high redshift $z \sim 5$,
the frequency-redshift and time-dilation 
effects are significant. 
\citet{Weiler02} noted that RSNe peak when the optical depth 
$\tau \sim 1$. Since the optical depth 
depends both on frequency and time 
with similar power index \citep{Weiler02},
the frequency-redshift and time-dilation effects
approximately cancel, so that a fixed observed frequency,
the peak time is nearly redshift-independent. 

As mentioned above, our luminosity function is likely biased 
toward the available bright events
in a small and incomplete sample.
To explore how this bias could affect our results, 
Fig.~\ref{fig:radio_obs_lum} shows how the detection rate
with $S_{\rm min}$ = 50 nJy at 1.4 GHz 
depends on core-collapse RSN luminosity.
Fig.~\ref{fig:radio_obs_lum}(a) shows that RSN with peak luminosities greater than
$10^{27} \ \rm erg \ s^{-1} \ Hz^{-1}$ contribute all of the detections
beyond redshift $z \sim 1$, and RSN need to peak brighter 
than $10^{28} \ \rm erg \ s^{-1} \ Hz^{-1}$ to be seen beyond $z \sim 3$.
Fig.~\ref{fig:radio_obs_lum}(b) similarly shows that the all-redshift detection
rate becomes substantial for explosions peaking 
$> 10^{26} \ \rm erg \ s^{-1} \ Hz^{-1}$.

\begin{figure}[!h]
\begin{center}
\includegraphics[width=0.6\textwidth]{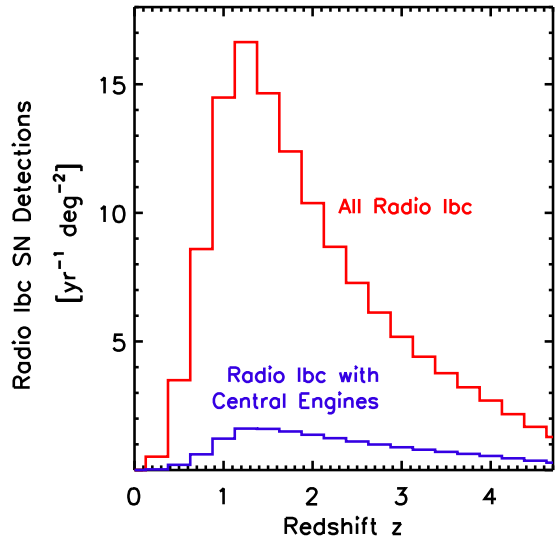}
\end{center}
\caption{
Predicted detection rate of Type Ibc supernovae
as a function of redshift.
In this plot we assume the sensitivity for SKA is $S_{\rm min}$ = 50 nJy.
The red curve shows all of the radio Ibc detections,
assuming $f_{\rm radio,Ibc} = 12\%$ \citep{Berger03}.
The blue curve shows only the detection rate for
Radio Ibc with central engines, assuming 0.5\% of
all of the Type Ibc RSNe are powered by central engines.
}
\label{fig:radio_Ibc}
\end{figure}

Type Ibc supernovae are of particular interest
given their association with long gamma-ray bursts
\citep[GRBs;][]{Galama98,Woosley93,Heger03}.
Fig.~\ref{fig:radio_Ibc} shows our predictions for 
Type Ibc detections of SKA per year per $\rm deg^2$ 
at 1.4 GHz 
with a survey sensitivity of 50 nJy. 
The red curve shows the radio Type Ibc detections, assuming
that Type Ibc represents 25\% of 
core-collapse events \citep{Li11b}, and 
$f_{\rm radio,Ibc} = 12\%$ with 
luminosity $\sim 10^{27} \ \rm erg \ s^{-1} \ Hz^{-1}$
\footnote{Here we simply assume a Gaussian distribution 
for the luminosity function centered 
at the specified luminosity with $\sigma = 1$.} \citep{Berger03}.
The blue curve shows the possible detections of the sub-class of Type Ibc 
supernovae that display extreme radio emission and hence might be powered by 
central engines and related to GRBs. 
We assume that 0.5\% of all Type Ibc supernovae 
are powered by central engines and have luminosities of
$\sim 10^{29} \ \rm erg \ s^{-1} \ Hz^{-1}$ \citep{Berger03}.
We adopted the spectrum of SN 1998bw, which is a Type Ic supernova 
\citep{Weiler02}.
Under these assumptions the SKA will be able to make
unbiased, untargeted detections of
$\sim 130$ radio Type Ibc supernovae per year per $\rm deg^2$,
and $\sim 20$ Type Ibc supernovae that might be connected to GRBs. 

Finally, we turn to SKA precursors.
The EVLA\footnote{http://www.aoc.nrao.edu/evla}, 
a current leading-edge radio interferometer operating at centimeter
wavelengths, is anticipated to reach a 1-$\sigma$ rms noise of 
$\sigma_I \sim 1 \ \rm \mu Jy$ or 
less in 10 hours,
while SKA is expected to reach $\sigma_I \sim$ 50 nJy in 10 hours. With data 
accumulated over repeated scans spanning over 1000 hours, 
an rms $\sigma_I \sim$ 5 nJy 
may be reached.
In synoptic surveys, we would expect EVLA 
to detect core-collapse events at a rate
$\sim 160 \ \rm RSNe \ yr^{-1} \ deg^{-2}$
over a redshift range $z = 0.5$ to 3
(Fig.~\ref{fig:radio_obs}).
A sample of this size over this redshift range will already
mark a major advance in the study of cosmic RSNe,
and further motivate the full SKA.
ASKAP and MeerKAT are expected to have sensitivities 
comparable to that of EVLA 
\citep{Johnston09, deBlok10},
hence we would expect these to detect RSNe 
with similar rates and redshift reach.

\subsection{Type Ia Supernovae}
\label{sect:Ia_predict}

If all Type Ia RSNe are dimmer than the weakest
limit presented in \S \ref{sect:radioIa},
the expected SKA detection rate is essentially zero.
For example, if a typical Type Ia has a radio luminosity equal to 
the lowest published limit,
$L = 8.1 \times 10^{24} \ \rm erg \ s^{-1} \ Hz^{-1}$, 
this can be seen with a sensitivity $S_{\rm min} = 50$ nJy 
out to a 
luminosity distance $\sim 300$ Mpc ($z \sim 0.08$).
While $\sim 3900$ cosmic Ia events should
occur per year out to this distance over 
the entire sky, $\ll 1$ events are expected
in the SKA field of view.
More optimistically, imagine a typical Type Ia radio luminosity is
$L = 10^{26} \ \rm erg \ s^{-1} \ Hz^{-1}$, 
which is below $L = 4.2 \times 10^{26} \ \rm erg \ s^{-1} \ Hz^{-1}$, 
the highest published limit \citep{Panagia06};
here the luminosity distance increases to $\sim 1400$ Mpc
($z \sim 0.28$).
In this case, we find an SKA Type Ia detection rate
$\sim 0.5 \ \rm yr^{-1} \ deg^{-2}$, 
based on the local cosmic Type Ia rate derived from
SDSS-II optical data
\citep{Dilday10},
$S_{\rm min} = 50 \ \rm nJy$, and $f_{\rm radio} = 10\%$.\footnote{
This also is implied by Fig.~\ref{fig:radio_obs_lum},
which is for core-collapse events that have a higher cosmic rate.}
We see that even optimistically, we expect 
fewer than one event per SKA field-of-view per year.
Even with $f_{\rm radio} = 100\%$, the detection rate
is still only $\sim 5 \ \rm yr^{-1} \ deg^{-2}$.
Therefore we conclude that SKA will make
few, if any, blind detections of Type Ia supernovae.

Targeted radio observations to follow up from 
nearby optical detections will probably be the best way
to search for such events.
For example, we expect 10 Type Ia events/year in the LSST sky
within $\sim 60 \ \rm Mpc$ ($z \sim 0.015$). Type Ia (or core collapse!)  
events within this distance
observed with $S_{\rm min} = 50$ nJy,
would be detectable at luminosities 
$L \ga  3.0 \times 10^{23} \ \rm erg \ s^{-1} \ Hz^{-1}$. 
Amusingly, this is close to the radio luminosity of SN 1987A.  

\section{Radio Survey Recommendations}
\label{sect:recommend}

A key requirement for detecting
weak radio emission from CSM-supernovae 
interactions is 
improved radio interferometer sensitivity.
High angular resolution -- below an arcsecond at 1.4 GHz, \citep{Weiler04} 
-- is also required to avoid
natural confusion and to help identify supernovae 
against background galaxies. This is similar to 
the maximum EVLA angular resolution at 1.4 GHz.
For comparison, the maximum anticipated SKA baseline 
length of 3000 km, producing angular resolution of 0.014 arcsecond at 1.4 GHz, 
is sufficient to distinguish different galaxies
and also to resolve galaxies as extended sources
within the observable universe with rms confusion limit of $< \ 3 \ \rm nJy$ 
at 1.4 GHz \citep{Carilli04}.

A key science goal of the SKA is to detect transient radio sources, 
both known (e.g. pulsars, GRBs), 
and as-yet unknown. 
This requires sophisticated transient detection and 
classification algorithms very likely running commensally 
with other large surveys planned by the SKA, 
such as the HI spectroscopic survey and deep continuum fields. 
We assume here that SKA transient detection algorithms 
will encompass automated detection of RSNe. 
For example, current parameterized models 
\citep{Weiler86, Weiler90, Montes97, Chevalier82a, Chevalier82b} 
based on available data predict patterns of 
spectral index evolution characteristic of supernovae in general, 
and supernova sub-types in particular.
This information could be exploited for RSN detection,
even potentially against a background of unrelated source variability. 
Broad frequency coverage is important 
in this regard \citep{Weiler04}.

The SKA intrinsically is a high dynamic-range instrument, 
given the sensitivity implied by the large collecting area. 
The most demanding SKA science applications 
will require a dynamic range of $10^7$:1. 
The detection of faint RSNe will require 
a dynamic range that falls within that envelope.

Although the lightcurves of RSNe show great diversity,
the luminosities of core-collapse supernovae 
usually change much slower  
in radio than in optical.
RSN lightcurves typically evolve on
timescales of weeks to years;
a useful lightcurve compilation appears in
\citet{Stockdale07}.
Thus the minimum survey cadence (revisit periodicity)
need not be any more frequent than this.
Also, we have shown that core-collapse RSNe 
can be found out to high redshift
with surveys pushing down to $S_{\rm min} = 50 \ \rm nJy$.
For SKA this corresponds to about 100 hours of exposure,
in line with planned deep field exposures
which are part of the key science.
Thus, SKA as currently envisioned is well-suited
to core-collapse discovery.

On the other hand, SKA probably will not have 
sufficient survey sensitivity for a volumetric 
search for Type Ia events, based on our current knowledge 
of the cosmic Type Ia rate and the upper limits in their luminosities 
set by the non-detection of these events.
Follow-up observations from 
other wavelengths will likely be 
the best way to search for Type Ia RSNe.

The small volume of the local universe
will limit nearby {\em untargeted} SKA detections of 
low-redshift core-collapse RSNe.
We estimate only $\sim 2$ core-collapse RSN detections per year per square degree
within redshift $z \sim 0.5$ (assuming a 50 nJy sensitivity at 1.4 GHz
and $\fradio=\fradval$).
Unless SKA has large sky coverage comparable 
to those of optical surveys, it will be hard to get statistical information 
from such a small sample.
Therefore, 
{\em targeted} radio followup of 
optically-confirmed nearby supernovae 
will be crucial to build a core-collapse RSN ``training set'' database
needed for refining
automatic identification and classification techniques.

With detection methods optimized based on low-$z$ radio data for
optically-identified events,
radio surveys can then be used
to independently detect core-collapse RSNe 
at high redshift based only on their radio emission.
As shown in Fig.~\ref{fig:radio_obs}, 
supernova searches at high redshift ($z \gtrsim 1$) 
will largely rely on radio synoptic surveys,
the inverse of the strategy proposed above for low-redshift domain.
Surveys for core-collapse RSNe will likely not be 
synoptic all-sky surveys due to
operational limitations, 
but will likely proceed in a limited set of sub-fields, 
visited over an hierarchical set of cadences to 
cover a range of time-scales for
general transient phenomena
and multiple commensal science objectives.
It is also important to match core-collapse RSNe survey sky coverage 
and cadence  
to that used in complementary optical surveys.
LSST will repeatedly scan the whole sky
every $\sim 3$ days. Thus
a cadence $\sim 1$ week
for RSNe sub-fields will be preferred for an SKA core-collapse supernova survey.

\section{Discussion and Conclusions}
\label{sect:science}

SKA's capability for unbiased synoptic searches over
large fields of view will revolutionize the discovery of
radio transients in general and core-collapse RSNe in particular
\citep{Gal-Yam06}.
The unprecedented sensitivity of SKA
could allow detection of core-collapse RSNe out to 
a redshift $z \sim 5$.
These detections will be {\em unbiased} 
and {\em automatic} in that
they can occur anywhere in the large SKA field of view without
need for targeting based on prior detection at other wavelengths.
With SKA, the core-collapse RSN inventory should increase 
from the current number of several dozen to
$ \sim 620 \ \rm yr^{-1} \ deg^{-2}$.
EVLA should detect 
$ \sim 160 \ \rm yr^{-1} \ deg^{-2}$,
and other SKA precursors should reap similarly large RSN
harvests.
In contrast, intrinsically dim RSNe such as 
Type Ia events and 1987A-like core-collapse
explosions are unlikely to be found blindly. 
However, the SKA (and precursor) sensitivities
will offer the possibility of detecting these events via 
{\em targeted} followup
of discoveries by optical synoptic surveys such as LSST.

The science payoff of large-scale RSNe searches touches
many areas of astrophysics and cosmology.
We conclude with examples of
possible science applications with the new era of RSN survey.
However, the true potential of
untargeted radio search is very likely    
beyond what we mention.

Non-prompt RSN emission requires the presence of
circumstellar matter, so such surveys will
probe this material and the pre-supernova winds producing it.
For core-collapse supernovae, pre-supernova winds 
should depend on the metallicities of the progenitor stars
\citep{Leitherer92,Kudritzki00,Vink01,Mokiem07},
and should be weaker in metal-poor environments with lower
opacities in progenitor atmospheres.
This effect should lead to correlations between
RSN luminosity and host metallicity,
as well as an evolution of the RSN luminosity function 
towards lower values at higher redshifts.
For Type Ia supernovae, 
the mass-loss rate from the 
progenitors depends on the nature of the binary system, i.e.,
single or double degenerate \citep{Nomoto84,Iben84,Webbink84}. 
Radio detection of Type Ia supernovae
will probe the mass and density profile of the surrounding environment and 
hence be valuable for studying Type Ia physics
\citep{Eck95,Panagia06,Chomiuk11}.

Large-scale synoptic RSNe surveys will complement 
their optical counterparts.
While optical surveys such as LSST
will provide very large supernova statistics 
at $z \lesssim 1$, radio surveys will be crucial 
for detections at higher redshifts.
The nature and evolution of dust obscuration 
presents a major challenge
for optical supernova surveys and supernova cosmology. 
Current studies suggest dust obscuration increases rapidly
with redshift,
but uncertainties are large.
\citet{Mannucci07} estimate 
that optical surveys may miss $\sim 60\%$
of core-collapse supernovae and $\sim 35\%$ of Type Ia supernovae
at redshift $z \sim 2$.
RSN observations, in contrast, are essentially unaffected by dust.
Thus,
high-redshift supernovae  
could be detected at radio wavelengths but 
largely missed in counterpart optical searches.
Comparing supernova detections in both optical and radio
will provide a new and independent way to measure dust dependence on redshift.
In particular, SKA will be a powerful tool 
to directly detect supernovae 
in dust-obscured regions at large redshift, and
therefore offer what may be the only means to study 
the total supernovae rate, star-formation, and dust behavior in
these areas.

Additionally, 
radio surveys will reveal rare and exotic events.
For example, some
Type Ibc supernovae
are linked to long GRBs 
\citep[Galama et al.~1998; and see reviews in][]{Woosley06,Gehrels09},
probably via highly relativistic jets 
powered by central engines and 
will manifest themselves in extremely luminous radio emission
\citep{Woosley93, Iwamoto98, Li99, Woosley99, Heger03}.
Thus
one might expect radio surveys
to preferentially detect more Type Ibc supernovae than other supernova types.
An unbiased sample of Type Ibc RSNe will 
provide new information about the circumstellar environments
of these explosions and thus probe the mass-loss effects
believed to be crucial to the Ibc pre-explosion
evolutionary path 
\citep{Price02,Soderberg04,Soderberg06,Crockett07,Wellons11}; 
in addition, a large
sample of Ibc RSNe will allow systematic study of 
the differences, if any, between those which do an do not
host GRBs
\citep{Berger03,Soderberg-Nakar06,Soderberg07}.

Furthermore,
radio surveys give unique new insight into a possible 
class of massive star deaths via direct collapse 
into black holes, with powerful neutrino bursts but 
no electromagnetic emission
\citep{MacFadyen99,Fryer99,MacFadyen01,Heger03}. 
These ``invisible collapses''
can be probed by comparing supernovae detected 
electromagnetically and the diffuse background of cosmic supernova 
neutrinos \citep[and references therein]{lfb}.
By revealing dust-enshrouded SNe, radio surveys will
make this comparison robust by removing 
the degeneracy between truly invisible events
and those which are simply optically obscured. 
Indeed, direct collapse events without explosions but with
relativistic jets are 
candidates for GRB progenitors.
A comparison among RSNe,
optical supernovae, GRBs, 
and neutrino observations will provide important clues
to the physics of visible and invisible collapses, 
and their relation with GRBs.

We thus believe that a synoptic survey in radio wavelengths
will be crucial in many fields of astrophysics,
for it will bring the first complete and unbiased RSN sample
and systematically explore exotic radio transients.
SKA will be capable of performing 
such an untargeted survey with its unprecedented sensitivity.
Our knowledge of supernovae 
will thus be firmly extended into 
the radio and to high redshifts.

\acknowledgments
We thank Kurt Weiler, Christopher Stockdale, 
and Shunsaku Horiuchi for encouragement and
illuminating conversations.
We are also grateful to Joseph Lazio 
for helpful comments that have improved this paper.


\begin{thebibliography}{bs}

\bibitem[Baklanov et al.(2005)]{Baklanov05} 
Baklanov, P.~V., Blinnikov, S.~I., \& 
Pavlyuk, N.~N.\ 2005, Astronomy Letters, 31, 429 

\bibitem[Bazin et al.(2009)]{Bazin09}
Bazin, G., et al.\ 2009, \aap, 499, 653

\bibitem[Berger et al.(2003)]{Berger03} 
Berger, E., Kulkarni, S.~R., Frail, D.~A., \& Soderberg, A.~M.
\ 2003, \apj, 599, 408 

\bibitem[Botticella et al.(2008)]{Botticella08} 
Botticella, M.~T., et al.\ 2008, \aap, 479, 49 

\bibitem[Cappellaro et al.(1999)]{Cappellaro99} 
Cappellaro, E., Evans, R., \& Turatto, M.\ 1999, \aap, 351, 459 

\bibitem[Cappellaro et al.(2005)]{Cappellaro05} 
Cappellaro, E., et al.\ 2005, \aap, 430, 83 

\bibitem[Carilli \& Rawlings(2004)]{Carilli04} 
Carilli, C.~L., \& Rawlings, S.\ 2004, New Astronomy Review, 48, 979 

\bibitem[Chevalier(1982a)]{Chevalier82a} 
Chevalier, R.~A.\ 1982, \apj, 259, 302

\bibitem[Chevalier(1982b)]{Chevalier82b} 
Chevalier, R.~A.\ 1982, \apjl, 259, L85

\bibitem[Chevalier(1998)]{Chevalier98} 
Chevalier, R.~A.\ 1998, \apj, 499, 810 

\bibitem[Chomiuk et al.(2011)]{Chomiuk11} 
Chomiuk, L., Soderberg, A.~M., Chevalier, R., Badenes, C., \& Fransson, C.\ 
2011, Bulletin of the American Astronomical Society, 43, \#304.05

\bibitem[Crockett et al.(2007)]{Crockett07} 
Crockett, R.~M., et al.\ 2007, \mnras, 381, 835 

\bibitem[de Blok et al.(2010)]{deBlok10} 
de Blok, E.~W.~J.~G., 
Booth, R., Jonas, J., \& Fanaroff, B.\ 2010, 
ISKAF2010 Science Meeting

\bibitem[Dahlen et al.(2004)]{Dahlen04} 
Dahlen, T., et al.\ 2004, \apj, 613, 189 

\bibitem[Dahlen et al.(2008)]{Dahlen08}
T. Dahlen et al., The Extended HST Supernova
Survey: The Rates of Type Ia and CC SNe at
high-z, in Arcetri Supernova Rates Workshop, 2008,
\url{http://www.arcetri.astro.it/~filippo/snrate08/Home.html}

\bibitem[Dahlen et al.(2010)]{Dahlen10} 
Dahlen, T., Strolger, L., \& Riess, A.~G.\ 2010, 
Bulletin of the American Astronomical Society, 42, \#430.23 

\bibitem[Dilday et al.(2010)]{Dilday10} 
Dilday, B., et al.\ 
2010, \apj, 713, 1026 

\bibitem[Eck et al.(1995)]{Eck95} 
Eck, C.~R., Cowan, J.~J., Roberts, D.~A., Boffi, F.~R., \& 
Branch, D.\ 1995, \apjl, 451, L53 

\bibitem[Frieman et al.(2008)]{Frieman08} 
Frieman, J.~A., et al.\ 2008, \aj, 135, 338 

\bibitem[Fryer(1999)]{Fryer99} 
Fryer, C.~L.\ 1999, \apj, 522, 413

\bibitem[Galama et al.(1998)]{Galama98}
Galama, T.~J., et al.\ 1998, \nat, 395, 670

\bibitem[Gal-Yam et al.(2006)]{Gal-Yam06} 
Gal-Yam, A., et al.\ 2006, \apj, 639, 331 


\bibitem[Gehrels et al.(2009)]{Gehrels09} 
Gehrels, N., Ramirez-Ruiz, E., \& Fox, D.~B.\ 2009, \araa, 47, 567 

\bibitem[Heger et al.(2003)]{Heger03} 
Heger, A., Fryer, C.~L., Woosley, S.~E., Langer, N., \& Hartmann, D.~H.\ 2003, \apj, 591, 288
 
\bibitem[Hopkins \& Beacom(2006)]{Hopkins06} 
Hopkins, A.~M., \& Beacom, J.~F.\ 2006, \apj, 651, 142 

\bibitem[Horiuchi et al.(2009)]{Horiuchi09} 
Horiuchi, S., Beacom, J.~F., \& Dwek, E.\ 2009, \prd, 79, 083013 

\bibitem[Horiuchi et al.(2011)]{Horiuchi11} 
Horiuchi, S., Beacom, J.~F., Kochanek, C.~S., Prieto, J.~L., Stanek, K.~Z., 
\& Thompson, T.~A.\ 2011, arXiv:1102.1977 

\bibitem[Iben \& Tutukov(1984)]{Iben84} 
Iben, I., Jr., \& Tutukov, A.~V.\ 1984, \apjs, 54, 335 

\bibitem[Iwamoto et al.(1998)]{Iwamoto98} 
Iwamoto, K., et al.\ 1998, \nat, 395, 672 

\bibitem[Johnston et al.(2009)]{Johnston09} 
Johnston, S., Feain, I.~J., \& Gupta, N.\ 2009, 
The Low-Frequency Radio Universe, 407, 446 

\bibitem[Kistler et al.(2008)]{Kistler08} 
Kistler, M.~D., Yuksel, H., Ando, S., Beacom, J.~F., 
\& Suzuki, Y.\ 2008, arXiv:0810.1959 

\bibitem[Komatsu et al.(2009)]{wmap5} 
Komatsu, E., et al.\ 2009, \apjs, 180, 330

\bibitem[Kudritzki \& Puls(2000)]{Kudritzki00} 
Kudritzki, R.-P., \& Puls, J.\ 2000, \araa, 38, 613 

\bibitem[Kulkarni et al.(1998)]{Kulkarni98} Kulkarni, S.~R., et 
al.\ 1998, \nat, 395, 663 

\bibitem[Leitherer et al.(1992)]{Leitherer92} 
Leitherer, C., Robert, C., \& Drissen, L.\ 1992, \apj, 401, 596 

\bibitem[Li et al.(2011a)]{Li11b} 
Li, W., et al.\ 2011, \mnras, 412, 1441 

\bibitem[Li et al.(2011b)]{Li11c} 
Li, W., Chornock, R., Leaman, J., Filippenko, A.~V., Poznanski, D., Wang, X., Ganeshalingam, M., \& Mannucci, F.\ 2011, \mnras, 412, 1473 

\bibitem[Li \& Chevalier(1999)]{Li99} 
Li, Z.-Y., \& Chevalier, R.~A.\ 1999, \apj, 526, 716 

\bibitem[Lien \& Fields(2009)]{lf} 
Lien, A., \& Fields, B.~D.\ 2009, 
Journal of Cosmology and Astro-Particle Physics, 1, 47 

\bibitem[Lien et al.(2010)]{lfb} 
Lien, A., Fields, B.~D., \& Beacom, J.~F.\ 2010, \prd, 81, 083001 

 \bibitem[LSST Science Collaborations et al.(2009)]{lsstsb} 
 LSST Science Collaborations, et al.\ 2009, LSST Science Book,
 arXiv:0912.0201 

\bibitem[MacFadyen \& Woosley(1999)]{MacFadyen99} 
MacFadyen, A.~I., \& Woosley, S.~E.\ 1999, \apj, 524, 262 

\bibitem[MacFadyen et al.(2001)]{MacFadyen01} 
MacFadyen, A.~I., Woosley, S.~E., \& Heger, A.\ 2001, \apj, 550, 410 

\bibitem[Madau et al.(1998)]{madau} 
Madau, P., della Valle, M., \& Panagia, N.\ 1998, \mnras, 297, L17 

\bibitem[Mannucci et al.(2007)]{Mannucci07}
Mannucci, F., Della Valle, M., \& Panagia, N.\ 2007, \mnras, 377, 1229

\bibitem[Mokiem et al.(2007)]{Mokiem07} 
Mokiem, M.~R., et al.\ 2007, \aap, 473, 603 

\bibitem[Montes et al.(1997)]{Montes97} Montes, M.~J., Weiler, 
K.~W., \& Panagia, N.\ 1997, \apj, 488, 792 

\bibitem[Munari et al.(1998)]{1997X}
Munari, U., Barbon, R., Piemonte, A., Tomasella, L.,
\& Rejkuba, M.\ 1998, \aap, 333, 159

\bibitem[Nakano \& Aoki(1997)]{1997ei} 
Nakano, S., \& Aoki, M.\ 1997, \iaucirc, 6795, 2 

\bibitem[Nomoto et al.(1984)]{Nomoto84} 
Nomoto, K., Thielemann, F.-K., \& Yokoi, K.\ 1984, \apj, 286, 644 

\bibitem[Palanque-Delabrouille et al.(2010)]{PD10}
Palanque-Delabrouille, N., et al.\ 2010, \aap, 514, A63

\bibitem[Panagia et al.(2006)]{Panagia06} 
Panagia, N., Van Dyk, S.~D., Weiler, K.~W., Sramek, R.~A., Stockdale, C.~J., 
\& Murata, K.~P.\ 2006, \apj, 646, 369 

\bibitem[Papenkova et al.(2001)]{Papenkova01}
Papenkova, M., Li, W.~D., Wray, J., Chleborad, C.~W., \& Schwartz, M.
\ 2001, \iaucirc, 7722, 1

\bibitem[Phillips(1993)]{Phillips93}
Phillips, M.~M.\ 1993, \apjl, 413, L105

\bibitem[Pooley et al.(2002)]{Pooley02} 
Pooley, D., et al.\ 2002, \apj, 572, 932 

\bibitem[Price et al.(2002)]{Price02} 
Price, P.~A., et al.\ 2002, \apjl, 572, L51 


\bibitem[Riess et al.(1996)]{Riess96}
Riess, A.~G., Press,
W.~H., \& Kirshner, R.~P.\ 1996, ApJ, 473, 88

\bibitem[Smartt et al.(2009)]{Smartt09} 
Smartt, S.~J., Eldridge, J.~J., Crockett, R.~M., 
\& Maund, J.~R.\ 2009, \mnras, 395, 1409 

\bibitem[SKA Design Reference Mission(2009)]{DRM09}
SKA Design Reference Mission, 2009, \\
{\tt \verb|http://www.skatelescope.org/PDF/DRM_v1.0.pdf|}.

\bibitem[Sako et al.(2008)]{Sako08}
Sako, M., et al.\ 2008, \aj, 135, 348

\bibitem[Salpeter(1955)]{Salpeter55} 
Salpeter, E.~E.\ 1955, \apj, 121, 161 

\bibitem[Soderberg et al.(2004)]{Soderberg04} 
Soderberg, A.~M., Frail, D.~A., \& Wieringa, M.~H.\ 2004, \apjl, 607, L13 

\bibitem[Soderberg et al.(2006)]{Soderberg-Nakar06} 
Soderberg, A.~M., Nakar, E., Berger, E., \& Kulkarni, S.~R.\ 
2006, \apj, 638, 930 

\bibitem[Soderberg et al.(2006)]{Soderberg06} 
Soderberg, A.~M., Chevalier, R.~A., Kulkarni, S.~R., \& 
Frail, D.~A.\ 2006, \apj, 651, 1005 

\bibitem[Soderberg(2007)]{Soderberg07} 
Soderberg, A.~M.\ 2007, 
Supernova 1987A: 20 Years After: Supernovae and Gamma-Ray Bursters, 937, 
492 

\bibitem[Stockdale et al.(2003)]{Stockdale03} 
Stockdale, C.~J., Weiler, K.~W., Van Dyk, S.~D., 
Montes, M.~J., Panagia, N., Sramek, R.~A., 
Perez-Torres, M.~A., \& Marcaide, J.~M.\ 2003, \apj, 592, 900 

\bibitem[Stockdale et al.(2006)]{rsnweb}
Stockdale, C. J., Kelley, M., Sramek, R.A., 
Williams, C. L., Van Dyk, S. D., Weiler, K. W., 
\& Panagia, N. 2006, New Radio Supernova Results.

\bibitem[Stockdale et al.(2007)]{Stockdale07} 
Stockdale, C.~J., Kelley, M.~T., Weiler, K.~W., 
Panagia, N., Sramek, R.~A., Marcaide, J.~M., Williams, C.~L.~M., 
\& van Dyk, S.~D.\ 2007, 
Supernova 1987A: 20 Years After: Supernovae and Gamma-Ray Bursters, 937, 264 

\bibitem[The LSST Collaboration(2007)]{lsst}
  The Large Synoptic Survey Telescope Collaboration 2007,
  Science Requirements Document, \\
  {\tt \verb|http://www.lsst.org/Science/docs/SRD.pdf|}

\bibitem[Tonry(2003)]{pan-starrs-sne}
J.~Tonry, (2003), Pan-STARRS Science Goals: Supernova Science, \\
\url{http://pan-starrs.ifa.hawaii.edu/project/science/precodr.html}; \\
\url{http://pan-starrs.ifa.hawaii.edu}.

\bibitem[Vink et al.(2001)]{Vink01} 
Vink, J.~S., de Koter, A., \& Lamers, H.~J.~G.~L.~M.\ 2001, \aap, 369, 574 

\bibitem[Webbink(1984)]{Webbink84} 
Webbink, R.~F.\ 1984, \apj, 277, 355 

\bibitem[Weiler et al.(1986)]{Weiler86} 
Weiler, K.~W., Sramek, R.~A., Panagia, N., van der Hulst, J.~M., 
\& Salvati, M.\ 1986, \apj, 301, 790 

\bibitem[Weiler et al.(1990)]{Weiler90} 
Weiler, K.~W., Panagia, N., \& Sramek, R.~A.\ 1990, \apj, 364, 611 

\bibitem[Weiler et al.(2002)]{Weiler02} 
Weiler, K.~W., Panagia, N., Montes, M.~J., 
\& Sramek, R.~A.\ 2002, \araa, 40, 387 

\bibitem[Weiler et al.(2004)]{Weiler04} 
Weiler, K.~W., van Dyk, S.~D., Sramek, R.~A., \& Panagia, N.
\ 2004, New Astronomy Review, 48, 1377 

\bibitem[Weiler et al.(2007)]{Weiler07} 
Weiler, K.~W., Williams, C.~L., Panagia, N., 
Stockdale, C.~J., Kelley, M.~T., Sramek, R.~A., Van 
Dyk, S.~D., \& Marcaide, J.~M.\ 2007, \apj, 671, 1959 

\bibitem[Weiler et al.(2009)]{Weiler09} 
Weiler, K.~W., Panagia, N., Sramek, R.~A., van Dyk, 
S.~D., Williams, C.~L., Stockdale, C.~J., \& Kelley, M.~T.
\ 2009, American Institute of Physics Conference Series, 1111, 440 

\bibitem[Wellons \& Soderberg(2011)]{Wellons11} 
Wellons, S., \& Soderberg, A.~M.\ 2011, 
Bulletin of the American Astronomical Society, 43, \#337.15 

\bibitem[Woosley(1993)]{Woosley93} 
Woosley, S.~E.\ 1993, \apj, 405, 273

\bibitem[Woosley et al.(1999)]{Woosley99} 
Woosley, S.~E., Eastman, R.~G., \& Schmidt, B.~P.\ 1999, \apj, 516, 788 

\bibitem[Woosley \& Bloom(2006)]{Woosley06} 
Woosley, S.~E., \& Bloom, J.~S.\ 2006, \araa, 44, 507 


\end{thebibliography}
\end{document}